\documentclass[letterpaper,prb,aps,twocolumn,showpacs]{revtex4}
\usepackage{graphicx,bm,color}    

\begin{document}
\title{Unconventional superconducting phases on a two-dimensional extended Hubbard model}
\author{Wen-Min Huang$^1$, Chen-Yen Lai$^1$, Chuntai Shi$^{2}$ and Shan-Wen Tsai$^1$}
\affiliation{$^1$Department of Physics and Astronomy, University of California, Riverside, CA 92521, USA.\\
$^2$Department of Physics and Astronomy, University of California, Irvine, CA 92697, USA.}

\begin{abstract} 
We study the phase diagram of the extended Hubbard model on a two-dimensional square lattice, including on-site ($U$) and nearest-neighbor ($V$) interactions, at weak couplings.
We show that the charge-density-wave phase that is known to occur at half-filling when $4 V > U$ gives way to a $d_{xy}$-wave superconducting instability away from half-filling, when the Fermi surface is not perfectly nested, and for sufficiently large repulsive $V$ and a range of on-site repulsive interaction $U$.
%
In addition, 
when nesting is further suppressed and in presence of a nearest-neighbor attraction, a triplet time-reversal breaking $(p_x+ip_y)$-wave pairing instability emerges, competing with the $d_{x^2+y^2}$ pairing state that is known to dominate at fillings just slightly away from half. 
At even smaller fillings, where the Fermi surface no longer presents any nesting,  
the $(p_x+ip_y)$-wave superconducting phase dominates in the whole regime of on-site repulsions and nearest-neighbor attractions, while $d_{xy}$-pairing occurs in the presence of on-site attraction. 
Our results suggest that zero-energy Majorana fermions can be realized on a square lattice in the presence of a magnetic field. For a system of cold fermionic atoms on a two-dimensional square optical lattice, both an on-site repulsion and a nearest-neighbor attraction would be required, in addition to rotation of the system to create vortices. We discuss possible ways of experimentally engineering the required interaction terms in a cold atom system.
\end{abstract}

\date{\today}


\maketitle

\section{Introduction}

An extended Hubbard model is generally employed as a theoretical framework of screened electronic interactions and regarded as a prototypical scenario for rich quantum phases in condensed matter physics\cite{Campbell,Yamaji,Weiss}. In a one-dimensional chain, for instance, an extended Hubbard model, including an on-site and nearest-neighbor interactions, presents correlated 
phases associated with the ratio of the two interactions\cite{Hirsch,Nakamura}. Recent indentification of the bond-charge-density-wave instability between charge and spin density-wave phases, at weak and strong interactions, completes the phase diagram of this model\cite{Furusaki,Sengupta,Sandvik,Zhang,Tam,Nishimoto}. Meanwhile, in a two-leg ladder, a checkerboard charge-ordered state has been proposed for all fillings between quarter and half, with on-site and nearest-neighbor repulsion\cite{Vojta}. Its application to the coupled quarter-filled ladders with coupling to the lattice has recently been studied\cite{Noack} to explain the spin gaps in the NaV$_2$O$_5$ material\cite{Ueda,Lemmens}.   

On a two-dimensional lattice, the extended Hubbard model has been considered a paradigmatic model to search for possible unconventional superconducting phases since the discovery of high-temperature superconductivity in the cuprates\cite{Scalapino}. Although a nearest-neighbor repulsion between electrons suppresses non-s-wave pairings tendencies\cite{Kabanov}, it is generally believed that the nesting of the Fermi surface plays a key role in driving 
unconventional pairing under purely repulsive interactions at weak couplings\cite{SC}. In the proximity of density-wave order, for instance, a chiral $d$-wave state has been found for an extended Hubbard model on both triangular and honeycomb lattices\cite{Honerkamp08,Kiesel131,Kiesel121}. Furthermore, following the recent experimental realization of a two-dimensional Kagome lattice for ultracold atoms\cite{Jo}, the phase diagram of the extended Hubbard model on a Kagome lattice has been established in the vicinity of van Hove fillings\cite{Wang}. It was shown that a possible $p$-wave charge and spin bond order can be triggered in the presence of a nearest-neighbor repulsion for van Hove fillings, then giving way to a $f$-wave superconducting phase when slightly doped away\cite{Kiesel122}. 

It has recently been proposed that under a nearest-neighbor attraction and on-site repulsion a singlet $(p+ip)$-wave pairing emerges on a honeycomb lattice\cite{Uchoa}. Along with the result that Majorana fermions can be generated as a zero-energy mode in the  excitation spectrum of a half-quantum vortex in a $(p+ip)$-wave superconductor\cite{Ivanov}, this indicates the possibility of creating  Majorana fermions in graphene in the presence of a magnetic field. However, a functional renormalization group study has shown that 
for a honeycomb structure, the $f$-wave pairing is preferred and is stabilized by introducing a next-nearest-neighbor attraction\cite{Honerkamp08}. Although triplet $p$-wave superconductivity has been proposed for an extended Hubbard model on a square lattice for purely repulsive interactions\cite{Hlubina}, the physics of long-range interactions, with possibly terms leading to competing instabilities, has been studied recently, leading that the existence of $p$-wave pairing is still an open question\cite{Onari,Raghu}. 

\begin{figure}[t]
\begin{center}
\includegraphics[width=7.8cm]{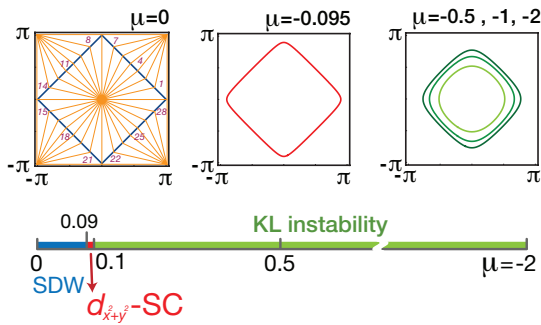}
\caption{(Color online) Fermi surfaces of a square lattice with nearest-neighbor hopping, at different chemical potentials $\mu$ (in units of $t$). Below, the phase diagram versus $\mu$ for $U=1$ and $V=0$. The Fermi surface patches used in this study are illustrated in the Fermi surface at $\mu=0$. }
\label{onsite}
\end{center}
\end{figure}

In this paper,  we study the phase diagram of an extended Hubbard model via a functional renormalization group (fRG) approach\cite{Shankar,Schonhammer}, including on-site $U$ and nearest-neighbor $V$ interactions, on a two-dimensional square lattice. The total Hamiltonian can be written as, $H=H_0+H_{\rm int}$, with the noninteraciting and interacting parts,
\begin{eqnarray}
&&H_0=-t\sum_{\langle ij\rangle,\alpha}\left(c^{\dag}_{i\alpha}c_{j\alpha}+{\rm H.c.}\right)-\mu \sum_{i} n_{i},\\
&&H_{\rm int}=U\sum_{i}n_{i\uparrow}n_{i\downarrow}+V\sum_{\langle ij\rangle} n_{i}n_{j},\label{int}
\end{eqnarray}
respectively, where $\langle ij\rangle$ represents nearest-neighbor pairs of sites, $n_{i}=n_{i\uparrow}+n_{i\downarrow}=\sum_{\alpha}c^{\dag}_{i\alpha}c_{i\alpha}$ and $\mu$ is the chemical potential. Without the nearest-neighbor interaction, the phase diagram in a fRG analysis is well developed in the limit of weak couplings\cite{Schonhammer}. For on-site attraction, $s$-wave superconductivity ($s$-SC) dominates at all fillings except for half-filling. The phase diagram for on-site repulsion, $U=1$, versus $\mu$ is sketched in Fig.~\ref{onsite}. In the vicinity of half-filling, the spin-density wave (SDW) dominates due to the strong nesting of the Fermi surface. With slight doping, $d_{x^2+y^2}$-SC emerges in a small regime of $\mu$ just away from half-filling. When the magnitude of $\mu$ is further increased, no instability develops up to the point where we stop the RG flows, at energy cutoffs lower than $10^{-6}t$. Kohn-Luttinger (KL) instability\cite{KL}, with extremely low critical temperatures, is expected to occur in this regime.

In presence of a nearest-neighbor interaction $V$, the phase diagram is much richer. We find that $d_{xy}$- and $(p_x+ip_y)$-wave pairing superconducting states develop. In the proximity of the charge-density-wave (CDW) order, $d_{xy}$-wave pairing emerges from the CDW instability with $U>0$ and $V>0$. When the nesting of the Fermi surface is decreased, a time-reversal symmetry breaking $(p_x+ip_y)$-SC arises from the $d_{x^2+y^2}$-SC with a sufficient large nearest-neighbor attraction. When nesting is completely suppressed, $(p_x+ip_y)$-SC dominates in the whole regime of $U\geq0$ and $V<0$, and the $d_{xy}$-pairing is only triggered with the help of an on-site attraction. Using a symmetry argument, we show that appearance of $(p_x+ip_y)$-SC on a square lattice for $V<0$ is generic  and robust due to the underlaying lattice structure, and can be used to create a zero mode Majorana fermion in the presence of a magnetic field. 

This paper is organized as follows. In Sec. II, we study the phase diagrams at and near half-filling ($\mu=0$ and $-0.095$). We will show that our fRG results are consistent with the previous studies at the half-filling. In Sec. III, we show the phase diagrams for chemical potential $\mu=-0.5$, $-1$ and $-2$ and study the possible unconventional SC states driven by the nearest-neighbor interaction. We discuss our results and conclusions in Sec. IV. 


\section{At and near half filling}

\begin{figure}[!t]
\begin{center}
\includegraphics[width=8.2cm]{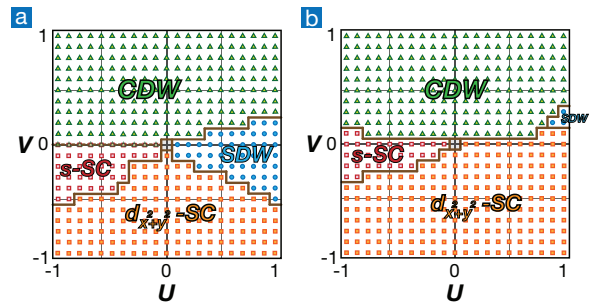}
\caption{(Color online) The phase diagram (a) at half-filling $\mu=0$, (b) at $\mu=-0.095$, parameterized by on-site $U$ and nearest-neighbor interaction $V$. The interaction terms $U$ and $V$ are in units of $t$ throughout this paper. }
\label{nesting}
\end{center}
\end{figure}

Starting with the bare Hamiltonian, Eq.~(\ref{int}), we follow standard fRG procedure integrating out high-energy modes, decreasing the energy cut-off $\Lambda$. The four-fermion terms in the resulting effective Hamiltonian are written in the form $g({\bm k}_1,{\bm k}_2,{\bm k}_3,\Lambda)\psi^{\dag}_{\alpha}({\bm k}_1)\psi^{\dag}_{\beta}({\bm k}_2)\psi_{\beta}({\bm k}_3)\psi_{\alpha}({\bm k}_1+{\bm k}_2-{\bm k}_3)$, in momentum space, with ${\bm k}=(k_x,k_y)$ and spin indices $\alpha,\beta$. In previous studies, the RG equations for models with $SU(2)$ and $U(1)$ symmetries have been systematically studied by Fermi surface discrete patch-approximation\cite{Schonhammer}. In this paper, to preserve the particle-hole symmetry of the non-interacting Hamiltonian at half-filling, all phase diagrams are obtained by the configuration of the Fermi surface patches illustrated in Fig.~\ref{onsite}. By integrating out high energy degrees of freedom and neglecting self-energy corrections, the RG flows of all couplings versus the decreasing running energy cutoff $\Lambda$ are computed. Before the system flows into the strong couplings regime, we truncate the RG process when the absolute magnitude of one of the couplings reaches $\sim30 t$. 

To determine the dominant instability, we decompose specific four-fermion interaction terms in the Hamiltonian as $\sum_{{\bm k},{\bm p}}\mathcal{V}_{\rm op}({\bm k},{\bm p},\Lambda)\hat{\mathcal{O}}^{\dag}_{\bm k}\hat{\mathcal{O}}_{\bm p}$, with $\hat{\mathcal{O}}_{\bm k}$ a bi-fermion operator for the order parameter (op) of SC, CDW, SDW or Pomeranchuk instability. Then, for a given order parameter channel, we further decompose, $\mathcal{V}_{\rm op}({\bm k},{\bm p},\Lambda)=\sum_{i}w_{\rm op}^i(\Lambda)f^{i*}_{\rm op}({\bm k},\Lambda)f^{i}_{\rm op}({\bm p},\Lambda)$, in normal modes, with $i$ a symmetry decomposition index. The leading instability can be determined by the most minimum eigenvalue $w_{\rm op}^{\rm min}(\Lambda)$ (largest magnitude), and the corresponding symmetry of the instability is given by the form factor $f^{\rm min}_{\rm op}(\bm{k})$\cite{FWang,Zanchi}.

At half-filling, the phase diagram has been studied extensively by several methods\cite{Kampf,Davoudi,Zhang}. In our fRG analysis, we include the regime of negative interactions and obtain the phase diagram in Fig.~\ref{nesting}a, parameterized by on-site interaction $U$ and nearest-neighbor interaction $V$ (in unit of $t$). 
The phase boundary between SDW and CDW is at $U\simeq4V$ for $U,V>0$, consistent with known results from previous studies. For the line of $V=0$ and $U<0$, we find that CDW and $s$-SC are degenerate, also in agreement with results in the literature\cite{YZhang}. However, the degeneracy is delicate and broken by introduction of a nearest-neighbor interaction: a slight nearest-neighbor attraction drives the system to a $s$-SC instability, instead of a CDW instability.   

We also find that 
for a sufficiently large nearest-neighbor attraction, a $d_{x^2+y^2}$-SC is triggered, even dominant over the $s$-SC in the regime of $U<0$. This $d$-wave pairing is linked with the nesting of the Fermi surface. In other words, if the nesting effect is suppressed, the $s$-SC is the dominant instability for a generic Fermi surface in the regime of $U<0$. Furthermore, without nesting effects, a $p$-wave, with lower angular momenta than $d$-wave, will eventually dominate in the regime of $V<0$ and $U>0$. However, negative $V$ combined with nesting, leads to $d_{x^2+y^2}$-SC.

By slightly doping away the half-filling, the $d_{x^2+y^2}$-SC overcomes the spin-density-wave instability that dominates for on-site repulsion. The phase diagram parameterized by $U$ and $V$ is illustrated in Fig.~\ref{nesting}b. The SDW is suppressed to the small regime between CDW and $d_{x^2+y^2}$-SC in the phase diagram. The degenerate line, $U<0$ and $V=0$, mentioned above, is dominated by $s$-SC instability since the Fermi surface is not perfectly nested. However, the CDW is still dominant in the overall regime of $V>0$. 

\begin{widetext}

\begin{figure}[t]
\centering
\includegraphics[width=17cm]{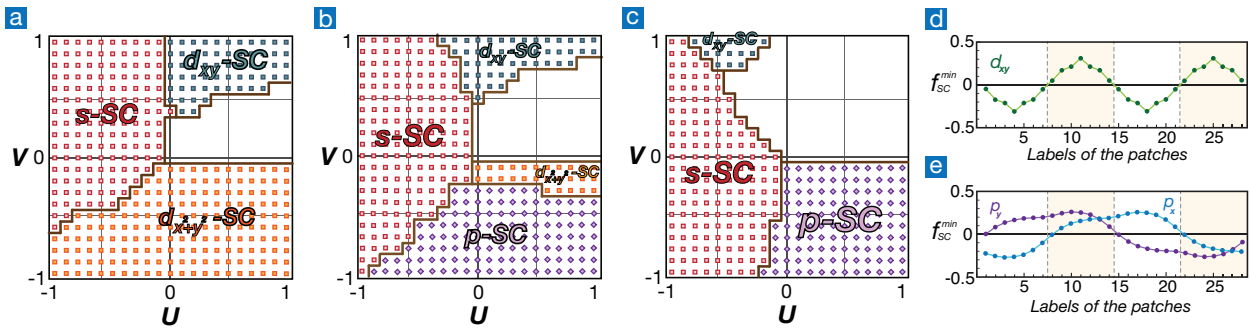}
\caption{(Color online) Phase diagrams for (a) $\mu=-0.5$, (b) $\mu=-1$ and (c) $\mu=-2$, parameterized by $U$ and $V$. The transparent regime means no instability is found before we stop the RG process, at which point the energy cutoff is lower than $10^{-6}t$. Form factors obtained from decoupling of pairing channels into bi-fermions during the fRG flows are illustrated for (d) $d_{xy}$-wave (e) $(p_x+ip_y)$-wave pairings. }
\label{doped}
\end{figure}

\end{widetext}


\section{doped systems}

Here, we increase doping, decreasing nesting of the until the density-wave instability no longer occurs. Then, as shown as Fig.~\ref{doped}a, the $d_{xy}$-wave SC arises from the CDW instability in the regime of $V>0$ and $U\geq0$. The form factors of the $d_{xy}$-SC in our fRG analysis is plotted in Fig.~\ref{doped}d. Although the $d_{xy}$-SC phase has been proposed for the purely repulsive models\cite{Hlubina,Onari,Raghu}, the $d_{xy}$-SC instability we find develops only with an appropriate nearest-neighbor repulsion and also tied with the proximity to nesting of the Fermi surface. It is only when nesting is barely suppressed so that CDW is no longer dominant, but CDW fluctuations are still expected to be strong, that this $d_{xy}$-SC phase emerges. 

As compared with the phase diagram at $\mu=-0.5$ in Fig.~\ref{doped}a, the $d_{xy}$-SC instability moves to the region of larger positive $V$ or negative $U$ when doping is increased, for $\mu=-1$, as shown in Fig.~\ref{doped}b. By looking at the RG flows of different couplings, we notice that the different behavior for the two fillings, $\mu=-0.5$ and $\mu=-1$, are mainly coming from the effect of the nesting vertices, that is, couplings $g({\bm k}_1,{\bm k}_2,{\bm k}_3,{\bm k}_4)$ with ${\bm k}_1+{\bm k}_2-{\bm k}_3-{\bm k}_4= (0,2\pi)$ or $(2\pi,0)$ or $(2\pi,2\pi)$. For $\mu=-1$, there are fewer nesting vertices flowing into large values as compared with the $\mu=-0.5$ case. Without the benefit of nesting vertices, the instability of $d_{xy}$- and $d_{x^2+y^2}$-SC are suppressed and eventually do not develop by the time we stop the RG flow. As a consequence, when the nesting is completely destroyed, in heavily doped cases, there is no instability found for purely repulsive interactions ($U>0$ and $V>0$), as shown in Fig.~\ref{doped}c.

The triplet $p$-SC phase tells an opposite story. In the regime of $V<0$, under the influence of Fermi surface nesting, $d_{x^2+y^2}$-SC is dominant, as shown as Fig.~\ref{doped}a. However, decreasing nesting suppresses the $d$-wave pairing and the $p$-SC emerges for large nearest-neighbor attraction, as shown in Fig.~\ref{doped}b. When the nesting is no longer present at all, $p$-SC dominates the regime of $V<0$ and $U\geq0$, as illustrated as Fig.~\ref{doped}c. The emergence of a triplet $p$-SC has a clear physical picture in a square lattice. In the presence of a nearest-neighbor attraction $V<0$, electronic pairings are triggered. To lower energy, a $s$-wave pairing with the lowest angular momenta would be favored. However, an on-site repulsion suppresses the $s$-wave pairing so that $p$-wave, with the second lowest angular momenta, is preferred.

Furthermore, the four-fold symmetry on a square lattice also indicates that the $p_x$- and $p_y$-wave instability channels must be degenerate. The degeneracy of two $p$-wave superconducting instabilities is indeed found in our fRG results and the associated form factors are plotted in Fig.~\ref{doped}e. Since the $p_x$- and $p_y$-wave superconducting states are degenerate, the gap functions will be given
as a linear combination of the two order parameters. By constructing the gap function as $\Delta_{\bm k}=\Delta_{p_x}({\bm k})+v\Delta_{p_y}({\bm k})$ with the complex coefficient $v$ containing a possible relative phase, the condensation energy in the standard BCS equation is obtained by the difference in energy between the superconducting and normal states, and is given by
\begin{eqnarray}\label{BCS}
\Delta E=E_{\rm SC}-E_{\rm N}=2\sum_{|\bm k|>k_F}\left[\epsilon_{\bm k}-\frac{2\epsilon_{\bm k}^2+|\Delta_{\bm k}|^2}{2\sqrt{\epsilon_{\bm k}^2+|\Delta_{\bm k}|^2}}\right],
\end{eqnarray}
where $\epsilon_{\bm k}$ is the dispersion relation of the non-interacting Hamiltonian. The second term of Eq.~(\ref{BCS}) is maximized when $v$ is purely imaginary, hence the time-reversal breaking pairing symmetry $p_x+ip_y$ is the energetically favored one\cite{Kiesel121,Kiesel131,Tsai,Platt}. Physically this is reasonable, since this choice of the phase guarantees that a gap forms everywhere along the Fermi surface, lowering the ground-state energy.

\section{discussion and conclusion}

In atomic Bose-Fermi mixtures, an effective attraction between fermions can be mediated by fluctuations of the Bose-Einstein condensate of the bosons\cite{Demler,Mathey,Suzuki}. In the presence of the mediated long-range attraction, the $p_x+ip_y$ wave superconducting state has been proposed in these systems\cite{Mathey}. However, the mechanism discussed here puts some constraint for the emergence of the $p_x+ip_y$-SC phase: in order to develop the $p_x+ip_y$-SC, long-range attraction and on-site repulsion is needed, as well as low density of fermions to avoid nesting of the Fermi surface, while large enough densities such that the Fermi energy is a large scale ($t > U, V$) to justify the validity of fRG results. Another possible way to manifest a mediated attraction is through another species of fermions in a Fermi-Fermi atom mixture\cite{WM}. By introducing an inter-species interaction in Fermi-Fermi mixtures, an effective interaction for one species of fermions can be obtained by tracing out the other species. This can be justified, for example, if the one of the species has a much smaller effective mass than the other. In this case\cite{WM2}, the mediated long-range interaction is found to decay rather rapidly and can be approximated by an effective on-site and nearest-neighbor interactions. By tracing out one species with low electronic density, an effective nearest-neighbor attraction can be obtained. Together with a bare hard-core on-site repulsion, it may provide the required conditions for the creation of the time-reversal breaking $p_x+ip_y$-wave pairing.

In conclusion, we study the phase diagram of an extended Hubbard model, including a on-site $U$ and nearest-neighbor $V$ interactions, on a two-dimensional square lattice. In the proximity of charge-density-wave order, the $d_{xy}$-SC overcomes the CDW, dominating in the regime of $U>0$ and $V>0$, from our fRG analysis. Accompanying the destruction of Fermi surface nesting, a time-reversal breaking $(p_x+ip_y)$-wave superconducting state arises in the regime of $V<0$. Our results indicates that, without nesting, the $(p_x+ip_y)$-SC on a square lattice under a nearest-neighbor attraction is the generic behavior due to the underlaying lattice structure, and can be used to create a zero mode Majorana fermion in the presence of a magnetic field.

\

\section*{ACKNOWLEDGMENT}

WMH sincerely acknowledges support from NSC Taiwan under Grant 101-2917-I-564-074, and computing clusters support from Pochung Chen in NTHU, Taiwan and TAPP in UCSC. CYL, CS, and SWT acknowledge support from NSF under Grant DMR-0847801 and from the UC-Lab FRP under Award number 09-LR-05-118602.

\end{document}